# Comprehensive Evaluations of Cone-beam CT dose in Image-guided Radiation Therapy via GPU-based Monte Carlo simulations

**Davide Montanari[1,2], Enrica Scolari[1,2], Chiara Silvestri[1,2], Yan Jiang Graves[1,2], Hao Yan[1,2], Laura Cervino[1,2], Roger Rice[2], Steve B. Jiang[1,2], Xun Jia[1,2]**

[1]Center for Advanced Radiotherapy Technologies, University of California San Diego, La Jolla, CA 92037-0843, USA
[2]Department of Radiation Medicine and Applied Sciences, University of California San Diego, La Jolla, CA 92037-0843, USA

E-mails: xunjia@ucsd.edu

Cone beam CT (CBCT) has been widely used for patient setup in image guided radiation therapy (IGRT). Radiation dose from CBCT scans has become a clinical concern. The purposes of this study are 1) to commission a GPU-based Monte Carlo (MC) dose calculation package gCTD for Varian On-Board Imaging (OBI) system and test the calculation accuracy, and 2) to quantitatively evaluate CBCT dose from the OBI system in typical IGRT scan protocols. We first conducted dose measurements in a water phantom. X-ray source model parameters used in gCTD are obtained through a commissioning process. gCTD accuracy is demonstrated by comparing calculations with measurements in water and in CTDI phantoms. 25 brain cancer patients are used to study dose in a standard-dose head protocol, and 25 prostate cancer patients are used to study dose in pelvis protocol and pelvis spotlight protocol. Mean dose to each organ is calculated. Mean dose to 2% voxels that have the highest dose is also computed to quantify the maximum. It is found that the mean dose value to an organ varies largely among patients. Moreover, dose distribution is highly non-homogeneous inside an organ. The maximum dose is found to be 1~3 times higher than the mean dose depending on the organ, and is up to 8 times higher for the entire body due to the very high dose region in bony structures. High computational efficiency has also been observed in our studies, such that MC dose calculation time is less than 5 min for a typical case.

## 1. Introduction

Cone-beam CT (CBCT) is one of the most widely used image guidance system for image guided radiation therapy (Jaffray *et al.*, 1999; Jaffray *et al.*, 2002; McBain *et al.*, 2006; Grills *et al.*, 2008). Despite the great benefits of accurate patient positioning, the widespread use of CBCT produce a considerable amount of excessive radiation dose to patients (Islam *et al.*, 2006; Wen *et al.*, 2007; Song *et al.*, 2008; Ding and Coffey, 2009). Even though the magnitude of the imaging dose per CBCT scan is small, especially compared to that of therapeutic dose, the cumulative dose over a long treatment course may not be negligible. Moreover, the imaging dose spreads over many critical organs, as opposed to the therapeutic dose that conforms well to the tumor area. These facts have raised a great concern about the associated radiation risks (Brenner *et al.*, 2003). It is hence of importance to quantitatively assess CBCT doses to patients in a comprehensive manner, in order to better understand the severity of this issue, and to evaluate and manage the associated risks (Murphy *et al.*, 2007; Alaei *et al.*, 2010).

Over the years, a lot of research efforts have been devoted to CBCT dose assessments. Among them, measurement-based methods utilize detectors, such as thermoluminescent dosimeters (TLDs) or ion chambers, to directly probe dose in CBCT scans. However, many of the researches were conducted in CTDI phantoms or other non-realistic phantoms (Islam *et al.*, 2006; Amer *et al.*, 2007; Song *et al.*, 2008). It is hence difficult to interpret the results in real patient cases, as the individual patient body geometry and compositions can greatly alter dose magnitudes and distributions. Although there are some measurements performed in anthropomorphic phantoms (Amer *et al.*, 2007; Palm *et al.*, 2010; Haelga *et al.*, 2012; Giaddui *et al.*, 2013), the use of phantoms limited measurements to only a few spatial locations. Because of the highly heterogeneous nature of a CBCT dose distribution due to the relatively low x-ray beam energy, the dose measurements at few isolated points may not capture some key information, e.g. maximum dose value, and may bias the results, e.g. mean organ dose.

Calculation-based approaches constitute another category of CBCT dose assessments. These methods compute dose distributions inside patients via certain calculation models. Of particular interest is Monte Carlo (MC) simulation. The calculation accuracy of this method has been well accepted due to its capability of modeling physical interactions between radiation and matter, as well as the system geometry (DeMarco *et al.*, 2004; DeMarco *et al.*, 2005; DeMarco *et al.*, 2007; Ding *et al.*, 2007; Ding and Coffey, 2009; Ding *et al.*, 2010; Li *et al.*, 2011b, a). Nonetheless, because of the statistical nature of the MC method, a large number of photons are required in a simulation to yield a desired level of precision. The associated long computation time hinders the wide application of this method for a systematic study of CBCT doses in patient groups.

Recently, computer graphics processing unit (GPU) has been employed to accelerate MC simulations for radiation dose calculations (Badal and Badano, 2009; Jia *et al.*, 2010; Hissoiny *et al.*, 2011; Jia *et al.*, 2011; Jahnke *et al.*, 2012; Jia *et al.*, 2012a; Jia *et al.*, 2012b). Particularly, x-ray photon transports in the kV energy range, e.g. in CT and CBCT, have been greatly sped up using the GPU platform (Badal and Badano, 2009; Jia *et al.*, 2012b). It was reported that it takes only minutes to conduct a CBCT dose calculation in a voxelized patient geometry using the gCTD package (Jia *et al.*, 2012b). This will greatly facilitate the comprehensive study of CBCT doses in real patient cases with a relatively large of sample size.

Yet, before conducting studies as such, it is necessary to validate the gCTD package in terms of dose calculation accuracy. Although the photon transport accuracy of gCTD has been previously established by comparing simulation results with those from EGSnrc (Kawrakow, 2000), the x-ray source model accuracy has not been investigated, which critically determines the realism of the generated photons before entering patient body and hence the overall dose calculation accuracy. Therefore, the first motivation of this paper is to commission the source model used in gCTD for the x-ray source in Varian On-Board-Imaging (OBI) system (Varian Medical Systems, Palo Alto, CA). The overall calculation accuracy will also be tested by comparing calculations with measurements.

The second motivation of this paper comes from the desire of comprehensively evaluating radiation dose from CBCT scans on a group of patients. It is expected that the dose distribution varies among patients to a large degree due to the different patient size, material, and relative locations between the patient and the x-ray source. Only when we study a large group of patients can we extract information to describe the CBCT dose in an objective manner. In this paper, CBCT dose distributions in three typical scan protocols of the OBI system will be calculated using the gCTD package. Relevant quantities will be computed to characterize the dose distributions, as well as inter- and intra-patient dose variations.

## 2. Methods and Materials

### 2.1 gCTD

gCTD is an MC dose calculation package for fast and accurate CT/CBCT dose calculations (Jia *et al.*, 2012b). It is developed on the GPU architecture under NVIDIA CUDA platform. gCTD transports photons in a voxelized patient geometry. Relevant photon physics in the diagnostic energy range, e.g. photoelectric effect, Rayleigh scatter, and Compton scatter, are included in the simulations. Electrons generated are not tracked and their energies are locally deposited. The implementations of gCTD are carefully tuned to achieve a high performance on the GPU platform. Specifically, Woodcock transport technique (Woodcock *et al.*, 1965) is employed to reduce the tedious voxel boundary crossing during the photon transport. A GPU-friendly method is developed to sample Rayleigh scattering angles and Compton scattering angles, reducing computational burdens in these two frequently preformed steps. The photon transport accuracy of gCTD has been validated against EGSnrc (Kawrakow, 2000), another popular MC package with well accepted accuracy. Meanwhile, it is found that high computational efficiency has been achieved due to the vast parallel processing capability of a GPU and the highly optimized algorithm and implementations. Imaging dose calculation for a Zubal head-and-neck phantom can be accomplished in ~17 sec with the average relative standard deviation of 0.4%.

130    Source model is an important component in gCTD, which critically determines the overall dose calculation accuracy in real patient cases. In gCTD, a point x-ray source is utilized. The source property is characterized by its energy spectrum and its photon fluence map, e.g. the relative photon fluence intensity as a function of photon direction. This is an effective source model, in that both of these two quantities are employed to describe photon distributions after exiting the whole source and before entering the

135    patient, regardless the detailed physical configurations inside the x-ray source. The validity of this model will be justified by the capability of successfully commissioning the model and using gCTD to compute dose in phantom cases. gCTD also supports realistic CBCT scan geometry, where the x-ray source rotates around the patient along a circular trajectory with defined source-to-axis distance. The start and the stop angles of

140    the scan, and the number of x-ray exposures are also controlled by the user.

*2.2 Commission of gCTD*

145    Before applying gCTD for dose calculations in patient cases, it is necessary to commission the x-ray beam model. In this study, we focus on three typical scan protocols of Varian OBI system on a Trilogy linear accelerator, that are routinely used in our clinic for patient setup purpose. These protocols are standard-dose head protocol, pelvis protocol, and pelvis spotlight protocol. The physical parameters of these protocols are listed in Table 1. Because of different bowtie filters and x-ray energies employed in these

150    protocols, each of them has to be commissioned individually following the procedures described below.

Table 1. Parameters of CBCT scan protocols in our studies.

| Protocol | Angular range (degree) | Bowtie filter | kVp | mAs per projection | Number of projections |
|---|---|---|---|---|---|
| Standard-dose head | 200 | Full-fan | 100 | 0.4 | 364 |
| Pelvis | 360 | Half-fan | 125 | 1.04 | 660 |
| Pelvis spotlight | 200 | Full-fan | 125 | 2.0 | 364 |

*2.2.1 Source fluence map and energy spectrum*

155    For the source fluence map, we conduct a CBCT air scan with no object between the source and the imager. The mAs level in these air scans are purposely reduced from the values listed in Table 1 to avoid saturation of the imager readings. This approach is valid, since modifying the mAs only changes the overall photon numbers hitting on the imager,

160    but not the spatial distribution. Hundreds of projection images are acquired in a scan.

These images are averaged to reduce noise level in the detected signals. The resulting image is regarded as the photon fluence map used in gCTD.

As for the energy spectrum, it is determined by comparing dose measurements and calculations in a water phantom. Specifically, we first conduct radiation dose
165 measurements in a physical water phantom using a PTW waterproof farmer chamber N30013, which is calibrated at the kV energy range. The x-ray source is fixed at one direction such that the beam impinges to the phantom normally. Tube mAs level of $I_c = 5$ mAs is used, which gives high enough readings while avoiding overheating the x-ray tube from repeated measurements. The source-to-surface distance (SSD) is set to
170 85 cm. The reduced SSD from a typical setting of 100 cm increases photon fluence and hence chamber readings. We acquire chamber reading $c(z)$ as a function of depth $z$ in water along the central axis of the beam.

Meanwhile, MC dose calculation is conducted using a homogenous water phantom of dimension $50.75 \times 50.75 \times 50.75$ cm$^3$ and the voxel size is $0.25 \times 0.25 \times 0.25$ cm$^3$.
175 The source location relative to the phantom is set identical to that in the experiment. In each run of the MC simulations for commissioning, $10^{10}$ source photons are simulated to yield results with a high precision. The photon fluence map obtained previously is used. The photon source spectrum is generated by a model developed by Boone et. al. (Boone and Seibert, 1997), with the kVp value set to the one used in the experiment. To
180 determine the source spectrum, we choose Al as the source filtration material and tune the filtration thickness, until a good match between the calculated depth dose curve and the measured chamber reading is observed. The choice of Al as the filtration material is because the x-ray tube in the OBI system has 2mm Al filtration and the bowtie filter is made of Al as well (Varian Medical Systems, 2008). It is expected that by varying the
185 thickness of Al filtration we could capture the major property of the source and hence get a good match between the simulation results and the measurements. Note that there is an overall scaling factor between the calculated results and the chamber reading due to different units in the two quantities, nC for the measurements $c(z)$ and keV/g per source photon for the MC simulation results $d_{MC}(z)$. The two data sets are first normalized by
190 their own mean values before making comparisons to remove the impacts of the different units.

### 2.2.2 Absolute dose calibration

195 Once the source spectrum and fluence map are obtained, we proceed to absolute dose calibration, which converts the MC calculated results into physical dose values in a unit of cGy. As such, we follow the procedure described in AAPM task group report 61 (Ma *et al.*, 2001) to convert ion chamber measurements $c(z)$ into absolute dose in water $d(z)$. In this process, correction factors are applied to correct raw chamber reading $c(z)$ for,
200 e.g. temperature and pressure, ion recombination, and chamber polarity. Dose in water is obtained by applying the mass energy absorption coefficient ratio of water to air $[\overline{\mu_{en}}/\rho]_{air}^{water}$, which is calculated using the listed mass energy absorption coefficient in the XCOM database (Berger *et al.*, 2010) and the previously obtained source energy spectrum. The measured dose value is further normalized by the mAs level used in experiments $I_c = 5$mAs, yielding dose per unit mAs, in a unit of cGy/mAs.

After that, a calibration factor $f$, which maps the MC calculation results into the actual dose values per mAs, is determined by using a least square fit, $\min |d(z) - f d_{MC}(z)|^2$. This has a closed form expression of $f = \langle d(z), d_{MC}(z) \rangle / \langle d_{MC}(z), d_{MC}(z) \rangle$, where $\langle \cdot, \cdot \rangle$ denotes vector inner product. The determination of this calibration factor $f$ utilizes dose measurements at all depths, as opposed to the depth of 2 cm suggested by the task group report 61, to reduce the impacts of inaccuracy associated in measurements and noise in MC simulations at a single point.

*2.3 Validation of gCTD*

*2.3.1 Water phantom*

We first validate the dose calculation in the water phantom. Specifically, we use a single x-ray projection with $I = 5$mAs and corresponding kVp value and bowtie filter in the CBCT protocol. We measure point dose using the ion chamber at depth $z = 2$ cm, and laterally 5 cm away from the central beam axis along the x and the y direction, where the coordinate system is illustrated in Figure 2(a). For the protocols using the full-fan bowtie filter, since the dose is expected to be approximately symmetric about the y axis, only dose at positive x side is measured. For the protocol with the half-fan bowtie filter, doses at both the positive and the negative x sides are measured. In all protocols, since dose does not vary significantly along the y direction, only one measurement at the positive y side is measured. The chamber readings are converted to dose in water following the standard procedure.

MC dose calculations in water are also conducted with $10^9$ source photon simulated. Calculated 3D dose distribution is converted to physical dose using the $f$ factor established previously and the $I = 5$mAs used in the experiment. Dose values at the measured locations are then extracted from the 3D dose array. The measured value $d_{Measure}$ and the MC calculated values $d_{MC}$ are compared, and relative error is reported as $e = |d_{MC} - d_{Measure}| / d_{Measure}$ to characterize the agreement between them.

*2.3.2 CTDI phantom*

We further validate gCTD dose calculations in a nested CTDI phantom (10 cm diameter cylinder phantom nested in a 16 cm diameter cylinder phantom) made of acrylic. Five readings are taken, one at the center, and four at periphery at radius $r = 4$ cm. The locations of each measurement are illustrated in Figure 1. For the pelvis mode, the 4 periphery measurements are expected to be the same due to rotational symmetry. Hence only one reading at the location number 2 is taken. The chamber readings are converted to dose in the acrylic medium using the mass energy absorption coefficient ratio with necessary chamber reading corrections.

As for the MC simulations, we created a digital phantom with the same size to the CTDI phantom. The material is set to acrylic and the density is its nominal value of $\rho = 1.19 \, \text{g/cm}^3$. MC dose calculations in this phantom are conducted using $10^9$ source photon. The calculated dose values at the measurement locations are compared with the experimental ones, and the relative errors are calculated.

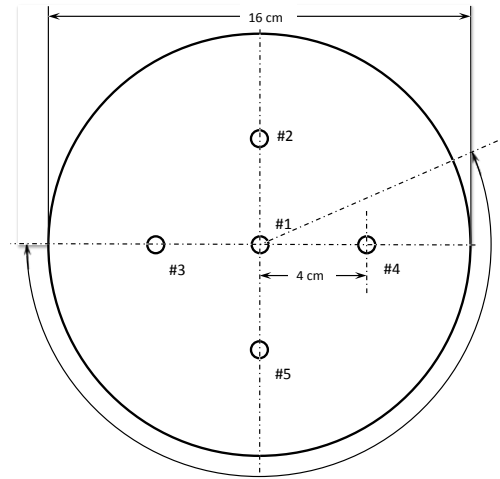

**Figure 1.** Illustrations of the CTDI phantom used in our experiments. Arrows indicate x-ray source range in the standard-dose head protocol and the pelvis spotlight protocol.

### 2.3.3 CBDI calculations

Third, we validate gCTD by calculating cone beam dose index (CBDI) and comparing the results with those published in literature. In this study, we created an acrylic CTDI phantom with 16 cm diameter for the head protocol, and with 32 cm diameter for the pelvis protocol. Dose calculations are conducted under the corresponding protocols, and dose values at the center, as well as four peripheral locations that are 1 cm away from the phantom boundary, are obtained. The measured central and the average of the four peripheral measurements are regarded as $CBDI_{100}^{center}$ and $CBDI_{100}^{periphey}$, respectively (Hyer and Hintenlang, 2010). The subscript 100 denotes an average of dose in a 100 mm range along the superior-inferior direction for each measurement. $CBDI_{100}^{w} = \frac{1}{3} CBDI_{100}^{center} + \frac{2}{3} CBDI_{100}^{periphey}$ is also calculated, which represents the average dose in the phantom. These calculated values are compared with those reported by Hyer et. al. (Hyer and Hintenlang, 2010), that are obtained by chamber measurements. Pelvis spotlight protocol is not studied here, since the values were not reported in that article.

### 2.3 CBCT dose in current clinical protocols

We proceed to evaluate CBCT dose in real patient cases, after establishing the calculation accuracy of gCTD. Two groups of patients are selected to investigate doses in the aforementioned three protocols. Specifically, 25 adult brain cancer patients were selected
to evaluate doses in the standard-dose head protocol. These patients were treated at our
institution with intensity-modulated radiation therapy (IMRT), and experienced CBCT-
based patient setup before treatments. Another 25 adult prostate cancer patients were also
selected to evaluate doses in the pelvis and the pelvis spotlight protocols. These patients
receive IMRT treatments and CBCT-guided patient positioning with the pelvis protocol
per the treatment protocol at our institution. However, we also compute doses from the
spotlight protocol on these patients as a hypothetical study regarding the doses in that
protocol.

For each of the patients, we first extract the CT image used in radiotherapy
treatment planning. The CT number at each voxel is converted into density value and
material type based on the calibration curve for our CT scanner. Treatment isocenter is
also obtained from the radiotherapy plans, which is assumed to be the isocenter of the
CBCT scans. After that, we conduct MC dose calculations using the source spectrum and
fluence map corresponding to the scan protocol. $10^9$ source photons are used to yield an
acceptable level of uncertainty. The resulting dose distribution is converted to the
absorbed dose per CBCT scan by multiplying with the $f$ factor and the total mAs level
for the scan.

Because of the available organ contours associated with those patients for IMRT
treatment planning purpose, we can evaluate organ doses based on the calculated 3D dose
distributions. As such, we first compute the mean dose $\bar{d}$ of an organ of interest by
simply averaging dose values over all voxels inside the organ. The mean dose value is
further averaged over all patients, $\langle\bar{d}\rangle$, to represent a typical dose to this organ from this
protocol. Because of the variations among patient size, anatomy, and relative locations of
CBCT isocenter to organs, dose to organs varies among patients. We characterize the
inter-patient dose variability by $\sigma = \Delta\bar{d}/\langle\bar{d}\rangle$, where $\Delta\bar{d}$ is the range of $\bar{d}$ among patients
for the organ of interest. The larger $\sigma$ is, the larger the dose varies among patients.
Meanwhile, it is also expected that the dose distribution inside an organ is not spatially
homogeneous. To characterize this fact, we first compute the mean dose to those 2% of
voxels inside the organ that receive the highest doses, denoted as $d_{2\%}$. This quantity
characterizes the maximum dose inside this organ to a certain extent. Note that we do not
use the maximum dose value directly, because it is the dose to a single voxel and may
have relatively large fluctuations in a MC calculation. $d_{2\%}$ is then averaged over all
patients, yielding $\langle d_{2\%}\rangle$. Finally, the intra-patient dose variability is represented by
$\eta = \langle d_{2\%}\rangle/\langle\bar{d}\rangle$. The higher $\eta$ is, the more dramatic the dose fluctuates spatially.

## 3. Results

### 3.1 gCTD calibration

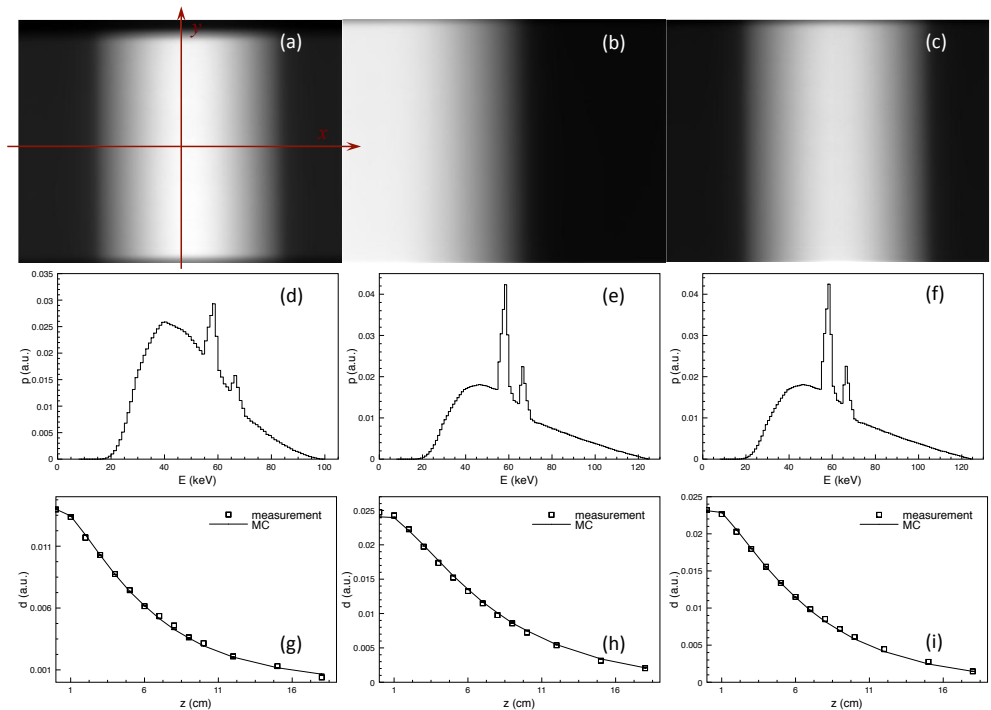

**Figure 2.** (a)~(c), fluence maps in head, pelvis, and pelvis spotlight protocols, respectively. (d)~(f) spectrums for the three protocols determined through the calibration process. (g)~(i) Comparisons of the calculated and the measured depth dose curves in the three protocols.

315    We first show in Figure 2 the commissioned effective source spectrums and the fluence maps for the three protocols. These results are obtained by tuning the Al filtration thickness to yield the best match between the calculated and the measured depth dose curves in water. The quality of this matching is also shown in Figure 2 (g)~(i).

For the standard-dose head protocol, the best fit between experiments and 320    calculations is achieved at 4.1 mm Al. In fact, the x-ray tube itself has 2 mm of equivalent Al filtration at the exit window (Varian Medical Systems, 2008). For the full-fan bowtie-filter used in this protocol, the thickness of the filter at the middle point is ~1.6 mm. The additional filtration observed in our calibration can be ascribed to other components along the beam path as well as source scattering, which effectively increases 325    the beam depth dose curve and hence the effective filtration thickness.

For the pelvis spotlight protocol, the effective filtration thickness is also found to be 4.1 mm Al. Comparing this protocol with the standard-dose head protocol, they have the same bowtie-filter but different kVp values. Because of the same source inherent filtration and the same bowtie-filter, it is reasonable that the same effective filtration 330    thickness is obtained in our commissioning. On the other hand, the effective filtration of the pelvis mode is 4 mm Al. This value is slightly different from those in the other two protocols, which could be due to the different bowtie-filter used in this mode.

Once the source spectrum and the fluence map is determined, we conducted absolute dose calibrations to determine the *f* factor that maps the MC calculated dose values into 335    the actual physical dose values. We do not list the *f* values here, as the values are only applicable to our MC-code and this calibration process. Hence, reporting the values does not have a general meaning.

### 3.2 gCTD validations

To demonstrate the calculation accuracy of gCTD after this commissioning process, we first conducted the measurements in a water phantom at depth $z = 2$ cm at different x and y coordinates. The calculated and measured dose values are listed in Table 2. The relative difference is maintained to be within a few percent, which demonstrates the acceptable accuracy achieved in this case.

Table 2. Comparison of the calculated and the measured doses in water.

| Protocol | Coordinate $(x, y)$ (cm) | Measured (mGy) | Calculated (mGy) | $e$ (%) |
|---|---|---|---|---|
| Standard-dose head | (5, 0) | 0.255 | 0.263 | 3.1 |
| | (0, 5) | 0.573 | 0.588 | 2.5 |
| Pelvis spotlight | (5, 0) | 0.489 | 0.497 | 1.6 |
| | (0, 5) | 1.002 | 1.039 | 3.7 |
| Pelvis | (-5, 0) | 1.101 | 1.174 | 6.6 |
| | (5, 0) | 0.856 | 0.921 | 7.6 |
| | (0, 5) | 1.108 | 1.117 | 3.2 |

For the validation study conducted in the CTDI phantom with 16 cm diameter, the results are summarized in Table 3. Again, the calculated and the measured doses are in good agreement, such that a few percent of relative dose difference is obtained.

Table 3. Comparison of the calculated and the measured doses CTDI phantom.

| Protocol | Location number | Measured (cGy) | Calculated (cGy) | $e$ (%) |
|---|---|---|---|---|
| Standard-dose head | 1 | 0.576 | 0.590 | 2.5 |
| | 2 | 0.376 | 0.367 | -2.3 |
| | 3 | 0.706 | 0.703 | -0.5 |
| | 4 | 0.531 | 0.559 | 5.2 |
| | 5 | 0.792 | 0.853 | 7.8 |
| Pelvis spotlight | 1 | 5.233 | 5.489 | 4.9 |
| | 2 | 3.239 | 3.434 | 6.0 |
| | 3 | 5.787 | 5.877 | 1.6 |
| | 4 | 5.202 | 5.528 | 6.3 |
| | 5 | 7.079 | 7.650 | 8.1 |
| Pelvis | 1 | 5.480 | 5.693 | 3.9 |
| | 2 | 5.822 | 5.619 | -3.5 |

Finally, Table 4 lists the results of CBDI values for the standard-dose head and the pelvis protocols. The reported CBDI values are taken from (Hyer and Hintenlang, 2010). Considering the uncertainty in measurements, the calculations and the measurements agree well.

355

Table 4. Comparison of the calculated and the measured CBDI values.

|  |  | Reported (mGy) | Calculated (mGy) | $e$ (%) |
|---|---|---|---|---|
| Standard-dose head | $CBDI_{100}^{center}$ | 5.34 | 5.43 | 1.7 |
|  | $CBDI_{100}^{periphery}$ | 5.09 | 5.85 | 14.9 |
|  | $CBDI_{100}^{w}$ | 5.17 | 5.71 | 10.4 |
| Pelvis | $CBDI_{100}^{center}$ | 14.38 | 16.87 | 17.3 |
|  | $CBDI_{100}^{periphery}$ | 25.17 | 26.26 | 4.3 |
|  | $CBDI_{100}^{w}$ | 21.57 | 23.13 | 7.1 |

### 3.3 Patient studies in current CBCT protocols

A typical dose distribution of the HN patient is shown in Figure 3. The dose distribution
360   is displayed in color scale overlaid with the patient CT image shown in gray scale. A few properties are apparent in these images. First, the dose to the eyes is low due to the $200°$ scan angle, in which the x-ray tube travels posteriorly to the patient head. Second, because of the low x-ray energy, the dose distribution is highly non-homogeneous. It decays fast as penetrating through the patient body. The inhomogeneity is exacerbated by
365   the high photoelectric effect cross section of bones compared to tissues, making the dose to bones a few times higher than to nearby tissues, as clearly indicated by the dose profile plots in Figure 4.

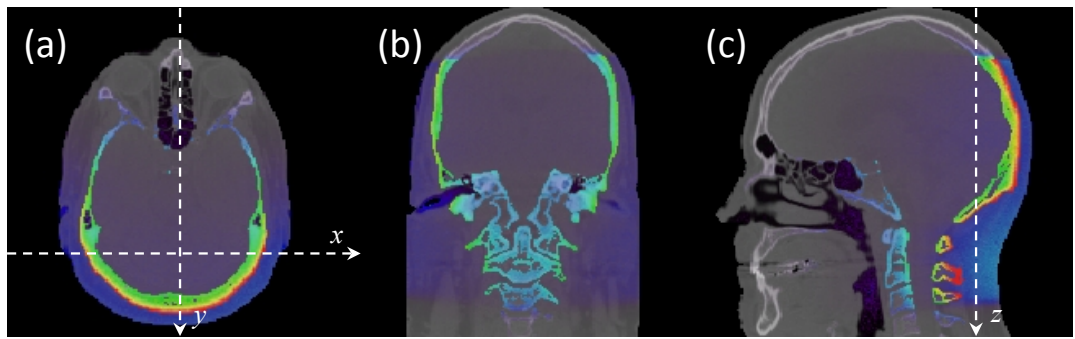

**Figure 3.** Dose distribution from a CBCT scan with the standard-dose head protocol. Dash lines in (a) are axes on which profiles in **Figure 4** are plotted.

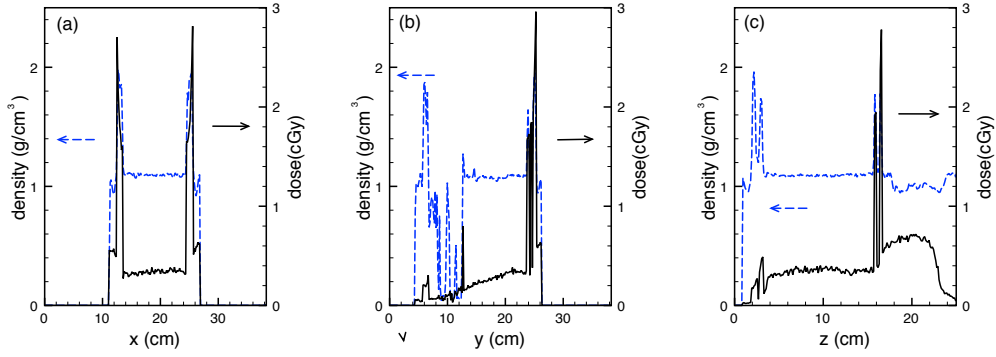

**Figure 4.** Dose profile (solid) and density profile (dash) along the axes shown in **Figure 3**.

370

Because of the availability of organ contours, we are able to compute mean dose per scan to these organs for the 25 patients we studied. Here, we report the mean organ dose $\langle \bar{d} \rangle$ averaged over all the patients, as well as the range of this value among all patients. The results are listed in Table 5. It is apparent that the dose level is low in a CBCT scan.

375   Meanwhile, $\sigma = \Delta\bar{d}/\langle\bar{d}\rangle$ is found to be 0.50~2.21, indicating that mean dose to an organ varies significantly among patients. As for the mean dose to those 2% of voxels with the highest doses inside each organ, it is much higher than the mean dose to that organ. The average value over all patients, as well as its range among patients are also listed in Table 5. We further compute the ratio of $\eta = \langle d_{2\%} \rangle / \langle \bar{d} \rangle$. It is found that $\eta$ ranges from 1.33 to

380   2.60 depending on organs. However, for the entire body, $\eta$ goes up to over 8, which can be ascribed to the higher dose to bony structures.

Table 5. Organ doses for the standard-dose head protocol.

| Organ | $\langle \bar{d} \rangle$ (cGy) | Range of $\bar{d}$ (cGy) | $\langle d_{2\%} \rangle$ (cGy) | Range of $d_{2\%}$ (cGy) | $\eta$ | $\sigma$ |
|---|---|---|---|---|---|---|
| Brain | 0.27 | 0.19~0.33 | 0.48 | 0.36~0.59 | 1.81 | 0.50 |
| Brainstem | 0.26 | 0.14~0.34 | 0.35 | 0.24~0.43 | 1.33 | 0.75 |
| Chiasm | 0.21 | 0.13~0.26 | 0.30 | 0.16~0.87 | 1.46 | 0.68 |
| Eyes | 0.093 | 0.03~0.24 | 0.13 | 0.06~0.31 | 1.39 | 2.21 |
| Optical nerves | 0.14 | 0.07~0.25 | 0.37 | 0.11~0.71 | 2.60 | 1.27 |
| Body | 0.33 | 0.20~0.47 | 2.74 | 2.11~3.34 | 8.34 | 0.83 |

For the pelvis and the pelvis spotlight protocols, we display the dose distributions

385   for a typical prostate cancer patient case in Figure 5 and the profiles along three axes in Figure 6. When comparing the two results, it is clear that the dose distribution in the spotlight protocol is more posteriorly located due to the x-ray source motion trajectory. This fact inevitably increases dose to rectum and reduces dose to bladder.

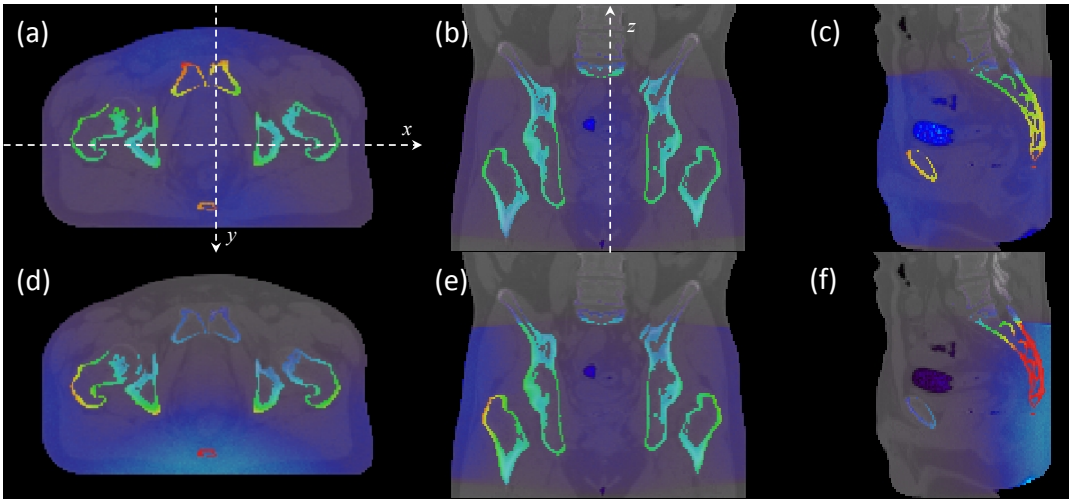

**Figure 5.** (a)~(c) dose distribution from a pelvis protocol. (e)~(f) dose distribution from a pelvis spot-light protocol.

390

The quantitative characterizations of dose to organs are listed in Table 6 and Table 7 for these two protocols, respectively. Again, the dose is highest in bony structures, e.g. femoral heads. As for the absolute dose comparisons between the two protocols, the average dose to rectum in the spotlight protocol is higher than that in the pelvis protocol,

395 caused by the higher mAs per projection in the spotlight protocol and the posteriorly located x-ray source. In contrast, the average dose to bladder in the pelvis protocol is higher. For the dose homogeneity inside each organ, the $\eta$ parameter indicates that dose distributions in most organs are not homogeneous. However, that of the rectum and of the bladder in the pelvis protocol is more homogeneous compared with other organs, which

400 can be ascribed to a large scan angle and the shallow organ locations. This fact can also be observed by visually comparing Figure 4 (a) and (d). The inter-patient dose variability is also found to be large, such that $\sigma$ can be up to 1.80.

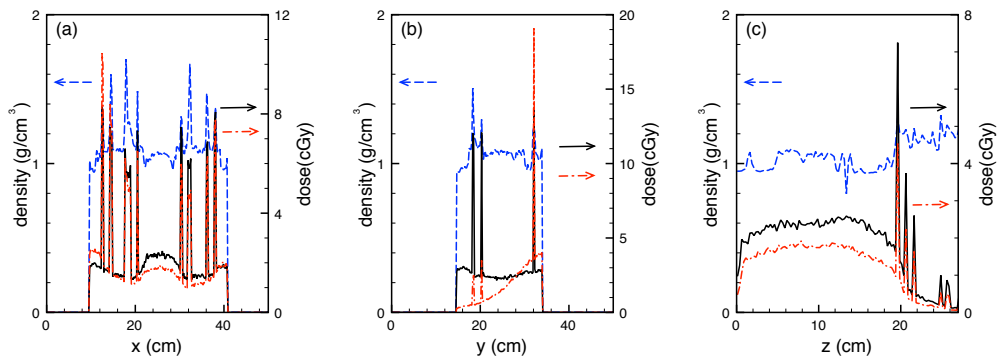

**Figure 6.** Dose profiles for the pelvis protocol (solid), those for the pelvis spotlight protocol (dash-dot), and density profiles (dash) along the axes shown in **Figure 5**.

405

Table 6.  Organ doses for the pelvis protocol.

| Organ | $\langle \bar{d} \rangle$ (cGy) | Range of $\bar{d}$ (cGy) | $\langle d_{2\%} \rangle$ (cGy) | Range of $d_{2\%}$ (cGy) | $\eta$ | $\sigma$ |
|---|---|---|---|---|---|---|
| Prostate | 2.04 | 1.16~3.06 | 4.69 | 2.36~10.52 | 2.04 | 0.94 |
| Rectum | 2.18 | 1.44~2.88 | 2.61 | 1.76~3.56 | 1.22 | 0.66 |
| Bladder | 2.13 | 1.15~3.15 | 2.82 | 1.42~4.10 | 1.23 | 0.94 |
| Femoral heads | 2.96 | 1.98~5.20 | 8.09 | 5.34~11.38 | 2.69 | 1.09 |
| Body | 1.21 | 0.66~2.08 | 6.16 | 3.13 | 4.69 | 1.17 |

*3.3 Computation time*

One of the advantages of using GPU-based MC dose engine is its fast processing speed. The dose calculation time obviously depends on patient size. In our study, all the patient cases have 256×256 voxels inside the transverse plane, and the voxel number along the superior-inferior direction ranges from 100 to 220. We have achieved a satisfactory speed. It takes less than 5 minutes on an Nvidia GTX 590 card to perform dose calculations with $10^9$ source photons for all the patient cases. The average relative uncertainties for those cases are found to be less than 1%. The reason why we simulate $10^9$ source photons is because we would like to have an accurate $d_{2\%}$ result, which is a quantity that is averaged over only a small number of voxels. It is hence sensitive to MC noise. If one were only interested in the mean dose to an organ, much fewer source photons would be sufficient, as the result is robust against MC uncertainty because of the average over many voxels. The computational time would also be shortened accordingly.

There are other steps in dose calculations, e.g. extracting doses to voxels inside organs of interest, and computing the mean dose and the average top 2% dose. However, compared to the MC step, the time spent on other steps is negligible.

Table 7.  Organ doses for the pelvis spotlight protocol.

| Organ | $\langle \bar{d} \rangle$ (cGy) | Range of $\bar{d}$ (cGy) | $\langle d_{2\%} \rangle$ (cGy) | Range of $d_{2\%}$ (cGy) | $\eta$ | $\sigma$ |
|---|---|---|---|---|---|---|
| Prostate | 1.79 | 1.06~2.55 | 3.87 | 2.28~5.81 | 2.16 | 0.83 |
| Rectum | 2.87 | 1.89~3.66 | 3.90 | 2.62~5.53 | 1.36 | 0.62 |
| Bladder | 1.09 | 0.41~2.36 | 1.97 | 0.71~3.36 | 1.81 | 1.80 |
| Femoral heads | 2.62 | 1.85~6.16 | 6.80 | 2.22~10.44 | 2.59 | 1.64 |
| Body | 1.16 | 0.61~2.02 | 6.96 | 4.57~11.59 | 6.00 | 1.22 |

**4. Conclusion and Discussions**

430     In this paper, we have commissioned a previously developed GPU-based MC package, gCTD, for CBCT imaging dose calculations. Source model parameters are obtained for three typically used CBCT protocols. The calculation accuracy is demonstrated by comparing the results with dose measurements. Through patient studies, doses to key organs in these three scan protocols are evaluated. It is found that the mean dose value to

435     an organ varies largely among patients. Moreover, dose distribution is highly non-homogeneous inside an organ. The maximum dose is found to be 1~3 times higher than the mean dose for those critical organs, and is up to 8 times higher for the entire body due to the presence of bony structures. High computational efficiency was also observed in our studies, such that MC dose calculation time is less than 5 min for a typical case with

440     $10^9$ source photons simulated.

     As for the clinical implications of the observed dose values, it is apparent that the doses reported in the three protocols are very small. The mean dose to an organ is of the order of mGy to cGy. However, whether or not these dose levels are of clinical concern will be subject to the specific context. If daily CBCT is repeatedly conducted on a patient

445     during a long treatment course, the total dose may not be negligible and may require specific managements (Alaei *et al.*, 2010). On the other hand, for a treatment course with only a few fractions, e.g. stereotactic radio-surgery, the imaging dose may be neglected, particularly considering the high therapeutic dose delivered per fraction. Other issues such as spatial inhomogeneity of the dose distributions may also call for attentions. The

450     maximum dose to an organ could be a clinical concern, although the high dose region is of a small area. Repeated CBCT scans will also likely smear the total dose distribution and reduce the maximum dose relative to the mean dose due to the patient position shift among different scans.

     We would also like to remark on the accuracy of our calculations. Even though the

455     accuracy of gCTD has been verified to a certain extent in the first half of this paper, the accuracy in real patient cases may also be limited by other practical factors, such as conversion from CT number to material properties. Compared to previous published CBCT dose data from measurements and from MC calculations, the organ dose values reported in this manuscript agrees with them sometimes only in terms of order of

460     magnitude, whereas the absolute quantities are different. Yet, this fact can be ascribed to a few factors. First, the measurement itself could have uncertainty to a certain degree. Second, the dose values highly depend on the location because of the spatial dose inhomogeneity. Third, the large inter-patient dose variability due to patient size and iso-center locations lead to vary patient-dependent results. In light of these facts, the doses

465     obtained in our studies should be accepted only semi-quantitatively, whereas the numbers should not be taken as exact values.

     Nonetheless, there are some features revealed by our studies, which should hold in general. Specifically, the observed large inter- and the intra- patient dose variability are expected to be true in general. The inter-patient organ-dose variability is mainly caused

470     by patient size and anatomy and relative locations of isocenter to organs, while the intra-patient variability is due to the low-energy x-ray and the anatomical inhomogeneity. Based on these findings, it may be necessary to report quantities other than the mean dose to an organ in order to objectively characterize the organ dose. This fact also demonstrates the necessity of patient-specific CBCT dose calculations, as proposed by many researches recently (Bacher *et al.*, 2005; Li *et al.*, 2011b, a; Jia *et al.*, 2012b).

475

Regarding the technical developments, our study demonstrated the feasibility of using the MC method to accurately and efficiently compute imaging dose distributions to patients. In particular, the short MC computation time is likely to be acceptable for integration into a clinical environment for patient-specific dose calculations. Nonetheless, we would also like to point out that automatic organ contouring remains to be a challenging problem, if organ-related dose quantities are desired. This will require further development in the future.

480

### Acknowledgements

485

This work is supported in part by Varian Master Research Agreement.

595